\documentstyle[12pt]{article}
\newcommand{\tr}[1]{\,{\rm tr}\,#1\,}

\begin{document}
\title{
\begin{flushright}
{\small SMI-01-96 }
\end{flushright}
\vspace{2cm}
Large $N$ Matrix Models
\\ and \\ $q$-Deformed Quantum Field Theories}
\author{
I.Ya.Aref'eva
\thanks
{ e-mail: arefeva@arevol.mian.su}
\\
Steklov Mathematical Institute,\\
Vavilov 42, GSP-1, 117966, Moscow, Russia}
\date {$~$}
\maketitle
\begin {abstract}
Recently it was shown that an  asymptotic behaviour of $SU(N)$ gauge theory
for large $N$ is described by  q-deformed  quantum field. The master fields
for large N theories satisfy to standard equations of relativistic field
theory but fields satisfy  $q$-deformed commutation relations with $q=0$.
These commutation relations are realized in the Boltzmannian Fock space.
The master field for gauge theory does not take values in a finite-dimensional
Lie algebra however there is a non-Abelian gauge symmetry. The gauge master
field for a subclass of planar diagrams, so called half-planar  diagrams,
is also considered. A recursive set of master fields summing up a matreoshka
of 2-particles reducible planar diagrams is briefly described.
\end {abstract}

\newpage
\section{Introduction}
\setcounter{equation}{0}
With great pleasure I dedicate my paper to Jurek Lukierski.
He made a great contribution in quantum field theory.
I hope that this paper will be interesting for him since his last
interests are related with q-deformed algebras \cite {Luk1,Luk2}.

In the last years there were considerations of models of quantum field theory
based on quantum groups and q-deformed commutation relations
\cite {Luk1}-\cite {APVV}.
The main problem with these approaches was how to find a physical
justification  for q-deformed  quantum field theory. Recently,
it was shown that there is a remarkable physical justification
for q-deformed  quantum field theory at least for $q=0$.
In  \cite {AVM} it was shown that such theory describes an asymptotic
behaviour of $SU(N)$ gauge theory for $N\to \infty$.

The large $N$ limit in QCD  where $N$ is the number of colours
enables us to understand qualitatively certain striking
phenomenological features of strong interactions \cite {tH}-\cite {AS}.
To perform an analytical investigation one needs to compute the sum of
all planar diagrams. Summation of planar diagrams has been performed only
in low dimensional space-time \cite {BIPZ,QCD2}.

It was suggested  \cite {Wit} that there exists
a master field which dominates the large $N$ limit.
There was an old problem in quantum field theory  how to construct
the master field for the large $N$ limit in QCD.
This problem has been discussed in many works, see for example
\cite {Haan}-\cite {Mig}.
More recently the problem has been reconsidered  \cite {GG}-\cite {AZsm}
by using methods of non-commutative (quantum)
probability theory \cite {Voi,Ac,ALV}.
Gopakumar and Gross \cite {GG} and Douglas \cite {Doug}
have described the master field using a knowledge of  all correlation
functions of a model.

Finally the problem of construction of the master field has been
solved in  \cite {AVM}.
It was shown that the master field
satisfies to standard equations of relativistic field theory but it
is quantized according to  $q$-deformed ($q=0$) relations
$$a(k)a^*(k')=\delta(k-k'),$$
where $a(k)$ and $a^*(k)$ are
annihilation and creation operators.
These operators have a  realization in the
free (Boltzmannian) Fock space. Therefore, to study the large N limit of QCD
it seems reasonable to develop methods of treatment of theory
in the Boltzmann space. Let us note that the fact that the master field
satisfies the same equations as usual relativistic fields makes actual
a development  of non-perturbative methods of investigation of these equations.

Quantum field theory in Boltzmannian Fock space has been considered in
\cite {AAV}-\cite {AZsm}.
Some special form of this theory realizes  the master field
 for a subset of  planar diagrams,
for the so called half-planar (HP) diagrams and gives an  analytical summations
of HP diagrams \cite {AAV}-\cite {AZsm}.
In this paper  a realization for the master field for
HP diagrams of gauge theories will be given. Then some recursive set of
master fields summing up  a more rich subclass of planar diagrams
will be also sketched. This subclass of diagrams contains a
matreoshka of 2-particles reducible planar diagrams.

In Section 2 consider the master field for  matrix models including
the master field for gauge field will be considered.
In Section 3  Boltzmann quantum field theory and
the HP master field for scalar and gauge theories  will be considered.

\section{Master Field for Planar Diagrams}
\setcounter{equation}{0}
\subsection {Zero-dimensional Master Field}

The master field $\Phi$ for the Gaussian
matrix model in zero dimensional space-time is defined by the relation
\begin {equation} 
                                                          \label {mc}
\lim _{N\to\infty} \frac{1}{Z_{N}}\int \frac{1}{N^{1+k/2}}\tr M^{k}
e^{-S(M)} dM =<0|\Phi ^{k}|0>
\end   {equation} 
for $k=1,2...$ , where the action $S(M)={1\over 2}\tr M^{2}$  and
$M$ is an Hermitian $N\times N $ matrix.

The operator $$\Phi= a+a^{*}$$  acts in the free or Boltzmannian Fock space
over C with vacuum vector $|0>$
\begin {equation} 
                                                          \label {vacuum}
a|0>=0
\end   {equation} 
and creation and annihilation operators
satisfying the relation
\begin {equation} 
                                                          \label {aa}
aa^{*}=1
\end   {equation} 

Let us recall that the free (or Boltzmannian)
Fock space $F(H)$ over the Hilbert space $H$
is  the tensor algebra over $H$
$${\cal T}(H)=\oplus_{n=0}^{\infty}H^{\otimes n}.$$
Creation and annihilation operators are defined as
$$a^{*}(f) f_{1}\otimes...\otimes f_{n}=f\otimes f_{1}\otimes...\otimes f_{n}$$
$$a(f) f_{1}\otimes...\otimes f_{n}=<f,f_{1}>\otimes f_{2}\otimes...\otimes
f_{n}$$
where $<f,g>$ is the inner product in $H$. One has
$$a(f)a^{*}(g)=<f,g>.$$
Here we consider the simplest case  $H=C$.

The relation  (\ref {mc}) has been obtained in physical
\cite {Haan} and mathematical \cite {Voi} works. It can be interpreted
as a central limit theorem in non-commutative (quantum) probability theory,
for a review see  \cite {ALV}.
The basic notion of non-commutative probability theory is an algebraic
probability space, i.e a pair $(A,h)$  where $A$ is an algebra
and $h$ is a positive   linear functional on $A$. An example of
the algebraic probability space is given by the algebra of random matrix
with (\ref {mc}) being    non-commutative central limit theorem.
As another example one can consider quantum groups.
Theory of quantum groups have received in the last years a lot of
attention  \cite {Dri}-\cite {Jimbo}. In this case
$A$ is the Hopf algebra of functions on the quantum group and $h$
is the quantum Haar measure.

The relations of theory of the master field
in the Boltzmannian Fock space with quantum groups was discussed
in  \cite {AAKV}.
There we defined the master field algebra and showed that
this algebra is isomorphic to the algebra of functions on the quantum
semigroup  $SU_{q}(2)$ for $q=0$.
In fact the master field algebra coincides with the algebra of the
so called central elements of the quantum group Hopf algebra.
Let us repeat the main steps of these observations.
In the Boltzmannian Fock space F(C) we have
\begin{equation}
                              \label {aca}
a^{*}a=1-|0><0|.
\end{equation}
 Let us define an operator
\begin{equation}\label {F}
F=e^{i\phi}|0><0|,
\end{equation}
where $\phi$ is an arbitrary real number. Then from
(\ref {vacuum}), (\ref {aca}), (\ref {aa}) one has the following
relations:
\begin {equation} 
                                                          \label {aF}
\begin{array}{ccc}
aF=0, & aF^{*}=0, & FF^{*}=F^{*}F, \\
aa^{*}=1,  &   a^{*}a+FF^{*}=1.
\end{array}
\end   {equation} 

We call the algebra (\ref {aF}) the  master field algebra.
>From equations (\ref {aF}) we get
$$(FF^{*})^{2}=FF^{*}$$
and the operator $FF^{*}$ is an orthogonal projector.

Now let us recall the definition of algebra of functions
$A_{q}=Fun(SU_{q}(2))$  on the quantum group  $SU_{q}(2)$.
The algebra $A_{q}$ is the Hopf algebra with generators
$a, a^{*}, c, c^{*}$ satisfying the relations

\begin {equation} 
                                                          \label {comcon}
\begin{array}{ccc}
ac^{*}=qc^{*}a,   & ac=qca,                 &  cc^{*}=c^{*}c, \\
a^{*}a+cc^{*}=1,  &  aa^{*}+q^{2}cc^{*}=1
\end{array}
\end   {equation} 
where $0<|q|<1$. Taking $q=0$ in  (\ref {comcon}) one gets the relations (\ref {aF})
if $F=c$. Therefore the master field algebra (\ref {aF})
is isomorphic to the algebra $A_{0}$ of functions
on the quantum (semi) group  $SU_{q}(2)$ for $q=0$.

$A_{q}$ is a Hopf algebra with the standard  coproduct,
$$\Delta :A_{q}\to A_{q}\otimes A_{q}$$
$$\Delta (g^{i}_{j})=\sum_{k=0,1}g^{i}_{k}\otimes g^{k}_{j}.$$
The Boltzman field$\Phi$
is a central element of the Hopf algebra $A_{q}$.

The bosonization of the quantum group $SU_{q}(2)$  \cite {APVV} gives
bosonization for the master field.
If $b$ and $b^{*}$ are the standard creation and annihilation operators
in the Bosonic Fock space,
$$[b,b^{*}]=1, ~~ b|0>=0,$$
then
\begin{equation}\label {ac}
a=\sqrt{\frac{1-q^{2(N+1)}}{N+1}}b, ~~ c=e^{i\phi}q^{N}
\end{equation}
satisfies the relations (\ref {comcon}). Here $N=b^{*}b, \phi$
is a real number. If $q\to 0$ one gets from (\ref {ac})
\begin{equation}                              \label {acq0}
a=\frac{1}{\sqrt{N+1}}b, ~~ c=e^{i\phi}|0><0|.
\end{equation}

Therefore the master field takes the form
\begin{equation}\label {Phi}
\Phi=b^{*}\frac{1}{\sqrt{N+1}}+\frac{1}{\sqrt{N+1}}b.
\end{equation}
The operator $N$ can be also written in terms of creation and annihilation
operators $a^+$, $a$, $N=\sum^\infty_{k=1}(a^+)^k(a)^k$.

\subsection{Master Field as a Classical Matrix}

Let us consider $U(N)$ invariant correlation functions for a model of
selfinteracting Hermitian scalar matrix field
$M (x)=(M_{ij} (x)),$ $ i,j=1,...,N$ in the D-dimensional
Euclidean space-time
\begin {equation} 
                                                          \label {icf}
<\tr (M(x_{1})...M(x_{k}))>=\frac{1}{Z_{N}}\int \tr (M(x_{1})...M(x_{k}))
e^{-S(M)} dM
\end   {equation} 
where $S(M)$ is the action $$S(M)=\int d^{D}x[{1\over 2}\tr (\partial M)^{2}
+\sum \frac{c_{i}}{N^{-1+i/2}}\tr M^{i}]$$  and
$M$ is an Hermitian $N\times N $ matrix,
\begin {equation} 
                                                          \label {nor}
Z_{N}=\int e^{-S(M)} dM
\end   {equation} 

Witten suggested  \cite {Wit} that there exists a master
field  which dominates in the large $N$ limit of  invariant correlation
functions of a matrix field, i.e.
\begin {equation} 
                                                          \label {Mc}
\lim _{N\to\infty} \frac{1}{Z_{N}}\int \frac{1}{N^{1+k/2}}
\tr (M(x_{1})...M(x_{k}))
e^{-S(M)} dM= \tr ( {\cal M}(x_{1})...{\cal M}(x_{k})),
\end   {equation} 
where ${\cal M}$ is some $\infty \times \infty$ matrix. Since
$\infty \times \infty$ matrix can be considered as an operator
acting in an infinite dimension space one can interpret the RHS of
(\ref {Mc}) as an expectation value of the product of some operators
$\Phi (x_{i})$
\begin {equation} 
                                                          \label {M}
\tr ({\cal M}(x_{1})...{\cal M}(x_{k}))
=<\Phi (x_{n}) ...\Phi (x_{1})>
\end   {equation} 
This interpretation gives an alternative definition of the master
field $\Phi (x)$ as a scalar operator which realizes the
following relation
\begin {equation} 
                                                          \label {icff}
\lim _{N\to\infty} \frac{1}{N^{1+k/2}}<\tr M(x_{1})...M(x_{k})>=
<\Phi (x_{n}) ...\Phi (x_{1})>
\end   {equation} 
where $<.>$ means some expectation value.
Therefore the problem is in constructing of a scalar field $\Phi$
acting in  some space so that the expectation value
of this field reproduces the large $N$ asymptotic of $U(N)$
invariant correlation functions of given matrix field.

\subsection{Free Master Field}

To construct master field for the  free matrix field let us calculate
the expectation value for free matrix field in the Euclidean space-time
\begin {equation} 
                                                          \label {4p}
\frac{1}{N^{3}}<\tr (M(x_{1})M(x_{2})M(x_{3})M(x_{4}))>^{(0)}=
\end   {equation} 
$$ \frac{1}{Z_{N}}\int \frac{1}{N^{3}}
\tr (M(x_{1})M(x_{2})M(x_{3})M(x_{4}))
e^{-S_{0}(M)} dM
$$
where the action $$S_{0}(M)=\int d^{D}x[{1\over 2}\tr (\partial M)^{2}].$$
We have
\begin {equation} 
                                                          \label {4pa}
\frac{1}{N^{3}}<\tr (M(x_{1})M(x_{2})M(x_{3})M(x_{4}))>^{(0)}=
D(x_{1}-x_{2})D(x_{3}-x_{4})+
\end   {equation} 
$$D(x_{1}-x_{4})D(x_{2}-x_{3})+\frac{1}{N}D(x_{1}-x_{3})D(x_{2}-x_{4}),
$$
here we use
\begin {equation} 
                                                          \label {pr}
<M_{ij}(x)M_{j'i'}(y)>^{(0)}=\delta _{ii'}\delta _{jj'} D(x-y),
\end   {equation} 
$D$ is an Euclidean propagator,
\begin {equation} 
                                                          \label {epr}
D(x-y)=\int \frac{d^Dk}{(2\pi ) ^D}\frac{e^{ik(x-y)}}{k^{2}+m^{2}}
\end   {equation} 
Let
\begin {equation} 
                                                          \label {phi}
\phi (x)=\phi ^+(x)+\phi ^-(x)
\end   {equation} 
be the Bolzmann field with creation and annihilation operators
satisfying the relations
\begin {equation} 
                                                          \label {0r}
\phi ^-(x) \phi ^+(y)=D(x-y),
\end   {equation} 
It is easy to check that
\begin {equation} 
                                                          \label {4r}
<0|\phi (x_{1})\phi (x_{2})\phi (x_{3})\phi (x_{4})|0>=
D(x_{1}-x_{2})D(x_{3}-x_{4})+D(x_{1}-x_{4})D(x_{2}-x_{3})
\end   {equation} 
where $|0>$ is a vacuum $\phi ^-(x)|0>=0=<0|\phi ^+(x)$.
The similar relation is true for an arbitrary n-point correlation function.
This consideration proves that the Euclidean Boltzmann field is a master
field for the Euclidean free matrix model.

Moreover, if we assume the relations
\begin {equation} 
                                                          \label {qr}
\phi _{q}^-(x) \phi _{q} ^+(y)+q\phi _{q}^+(y)\phi _{q}^-(x)=D(x-y),
\end   {equation} 
then we get
\begin {equation} 
                                                          \label {4rq}
<0|\phi _{q}(x_{1})\phi _{q}(x_{2})\phi _{q}(x_{3})\phi _{q}(x_{4})|0>=
\end   {equation} 
$$D(x_{1}-x_{2})D(x_{3}-x_{4})+D(x_{1}-x_{4})D(x_{2}-x_{3}) +
qD(x_{1}-x_{3})D(x_{2}-x_{4}),
$$
i.e. relation (\ref {4rq}) reproduces (\ref {4pa}) if we identify
\begin {equation} 
                                                          \label {Nq}
\frac{1}{N}=q.
\end   {equation} 

We can also consider the Minkowski space time.
To avoid misunderstanding we use a notation $M^{(in)}$ for the free Minkowski
matrix field. One has
\begin {equation} 
                                                          \label {1.14}
<0|M^{(in)}_{ij}(x)M^{(in)}_{pq} (y)|0>=\delta_{iq}\delta_{jp}
D^{-}(x-y)
\end   {equation} 
where
$$
D^{-}(x)=\frac{1}{(2\pi)^{3}}\int e^{ikx}\theta (-k^{0})
\delta (k^{2}-m^{2})dk.
$$

Let consider  the free scalar Boltzmannian field $\phi^{(in)}(x)$
given by
\begin {equation} 
                                                          \label {1.7}
\phi^{(in)}(x)=\frac{1}{(2\pi)^{3/2}}\int \frac{d^{3}k}
{\sqrt{2\omega (k)}}(a^*(k)e^{ikx}+a(k)e^{-ikx}) ,
\end   {equation} 
where $\omega (k)= \sqrt{k^{2}+m^{2}}$.  It satisfies to
the Klein-Gordon equation
$$(\Box + m^2)\phi^{(in)} (x)=0$$
and it is an operator in the Boltzmannian Fock space
with relations
\begin {equation} 
                                                          \label {1.8}
a(k)a^*(k')=\delta^{(3)}(k-k')
\end   {equation} 
and vacuum $|\Omega_0),~ ~ a(k)|\Omega_0)=0$. A systematical
consideration  of the Wightman formalism  for Boltzmannian
fields is presented in \cite {AAVMin}.

$a(k)$ and $a^*(k)$ act in  the Boltzmannian
Fock space $\Gamma (H)$ over  $H=L^2(R^3)$
and one uses notations such as $a(f)=\int a(k)f(k)dk$.
An $n$-particle state is created from the vacuum $|\Omega_{0})
=1$ by the usual formula
$$
|k_{1},...,k_{n})=a^{*}(k_{1})...a^{*}(k_{n}) |\Omega_{0})
$$
but it is not symmetric under permutation of $k_{i}$.

The the following
basic relation takes place
\begin {equation} 
                                                          \label {1.B}
\lim _{N \to \infty}
\frac{1}{N^{1+\frac{k}{2}}}<0|\tr(
(M^{(in)}(y_{1}))^{p_{1}}...(M^{(in)}(y_{r}))^{p_{r}})|0>
\end   {equation} 
$$
=(\Omega_{0}|(\phi^{(in)} (y_{1}))^{p_{1}}...(\phi^{(in)} (x_{r})
)^{p_{r}}|\Omega_{0})
$$
where $k=p_{1}+...+p_{r}$.
To prove  (\ref {1.B})  one uses the Wick theorem
for the Wightman functions and 't Hooft's graphs with double lines.
According to the Wick theorem we represent  the vacuum
expectation value in the L.H.S. of (\ref {1.B}) as a sum
of 't Hooft's graphs with the propagators   (\ref {1.14}).
Then in the limit  $N \to \infty $  only non-crossing
(rainbow) graphs are nonvanished. We get the same expression
if we compute the R.H.S. of (\ref {1.B}) by using the relations
(\ref {1.8}), i.e. by using the Boltzmannian Wick theorem.

\subsection{Master Field for Interacting Matrix Scalar Field}
To construct the master field for interacting quantum field theory
 \cite {AVM} we have to work in Minkowski
 space-time and use the Yang-Feldman formalism  \cite {YF}-\cite {Kal}.
Let us consider a model of an Hermitian scalar matrix field
$M (x)=(M_{ij} (x)),$ $ i,j=1,...,N$ in the 4-dimensional
Minkowski space-time with the field equations
\begin {equation} 
                                                          \label {1.1}
(\Box + m^{2})M (x)=J(x)
\end   {equation} 
We take the current  $J(x)$ equal to
\begin {equation} 
                                                          \label {1.2}
J(x)=-\frac{g}{N} M^3 (x)
\end   {equation} 
where $g$ is the coupling constant but one can take a more general
polynomial over $M(x)$.
One integrates eq (\ref {1.1}) to get the Yang-Feldman equation
\cite {YF,BD}
\begin {equation} 
                                                          \label {1.3}
M(x)=M^{(in)}(x)+\int D^{ret}(x-y)J(y)dy
\end   {equation} 
where $D^{ret}(x)$ is the retarded Green function for the Klein-Gordon
equation,
$$
D^{ret}(x)=
\frac{1}{(2\pi)^{4}}\int \frac{e^{-ikx}}{m^{2}-k^{2}-
i\epsilon k^{0}}dk
$$
and  $M^{(in)}(x)$ is a  free Bose field. The $U(N)$-invariant
Wightman functions  are defined as
\begin {equation} 
                                                          \label {1.4}
W(x_{1},...,x_{k})=
\frac{1}{N^{1+\frac{k}{2}}}<0|\tr(M(x_{1})...M(x_{k}))|0>
\end   {equation} 
where   $|0>$ is the Fock vacuum for the free field $M^{(in)}(x)$.

We will show that the limit of functions (\ref {1.4})  when
$N\to \infty$ can be expressed in terms of a quantum field $\phi (x)$
(the master field) which is a solution of the equation
\begin {equation} 
                                                          \label {1.5}
\phi(x)=\phi^{(in)}(x)+\int D^{ret}(x-y)j(y)dy
\end   {equation} 
where
\begin {equation} 
                                                          \label {1.6}
j(x)=-g \phi^3 (x)
\end   {equation} 
The master field  $\phi(x)$ does not have matrix indexes.

The following theorem is true.

{\bf Theorem 1.} {\it At every order of perturbation theory in the
coupling constant one has the following relation
\begin {equation} 
                                                          \label {1.10}
\lim _{N \to \infty}
\frac{1}{N^{1+\frac{k}{2}}}<0|\tr(M(x_1)...M(x_k))|0>
=(\Omega_{0}|\phi (x_{1})...\phi (x_{k})|\Omega_{0})
\end   {equation} 
where   the field $M (x)$ is defined by  (\ref {1.3})
and  $\phi (x)$ is defined by  (\ref {1.5})}.

The proof of the theorem see in \cite {AVM}.

\subsection{Gauge field}

In this section we construct the master field for  gauge field
theory. Let us consider the Lagrangian
\begin {equation} 
                                                          \label {3.1}
L=\tr\{-\frac{1}{4}F_{\mu \nu}^{2}-\frac{1}{2\alpha}
(\partial_{\mu}A_{\mu})^{2}+\bar{c}\partial_{\mu}
\nabla_{\mu}c\}
\end   {equation} 
where  $A_{\mu}$ is the gauge field for the $SU(N)$ group,
$c$ and $\bar{c}$ are the Faddeev-Popov ghost fields  and $\alpha$
is a gauge fixing parameter.  The fields $A_{\mu}$, $c$ and $\bar{c}$
take values in the adjoint representation. Here
$$
F_{\mu \nu}=\partial_{\mu}A_{\nu}-\partial_{\nu}A_{\mu}
+\frac{g}{N^{\frac{1}{2}}}[A_{\mu},A_{\nu}],~~~
\nabla_{\mu}c=\partial_{\mu}c+\frac{g}{N^{\frac{1}{2}}}[A_{\mu},c],
$$
$g$ is the coupling constant.
Equations of motion have the form
\begin {equation} 
                                                          \label {3.2'}
\nabla_{\mu} F_{\mu \nu} +
\frac{1}{\alpha}\partial_{\nu}\partial_{\mu}A_{\mu}+
\frac{g}{N^{\frac{1}{2}}}\partial_{\nu}\bar{c} c+
\frac{g}{N^{\frac{1}{2}}}c\partial_{\nu}\bar{c}=0,
\end   {equation} 
$$
\partial_{\mu}(\nabla_{\mu}c)=0,~~~
\nabla_{\mu}(\partial_{\mu}\bar{c})=0
$$
One writes these equations in the form
\begin {equation} 
                                                          \label {3.2}
\Box   A_{\nu} -(1-  \frac{1}{\alpha})\partial_{\nu}
\partial_{\mu}A_{\mu}=J_{\nu},
\end   {equation} 
$$
\Box c=J,~~~\Box \bar{c}=\bar{J},
$$
where
$$
J_{\nu} = -\frac{g}{N^{\frac{1}{2}}}\partial_{\mu}[A_{\mu},A_{\nu}]
-\frac{g}{N^{\frac{1}{2}}}[A_{\mu},F_{\mu\nu}]-
\frac{g}{N^{\frac{1}{2}}} \partial_{\nu} \bar{c}c-
\frac{g}{N^{\frac{1}{2}}}c\partial_{\nu}\bar{c},
$$
$$
  J = -\frac{g}{N^{\frac{1}{2}}} \partial_{\mu}[A_{\mu},c],~~~
\bar{J}=- \frac{g}{N^{\frac{1}{2}}}[A_{\mu},\partial_{\mu}\bar{c}]
$$
>From  (\ref {3.2}) one gets the Yang-Feldman equations
\begin {equation} 
                                                          \label {3.3}
A_{\mu}(x)=A_{\mu}^{(in)}(x)+  \int D^{ret}_{\mu\nu}
(x-y)J_{\nu}(y)dy,
\end   {equation} 
$$c(x)=c^{(in)}(x)+  \int D^{ret}(x-y)J(y)dy,
~ ~\bar{c}(x)=\bar{c}^{(in)}(x)+
\int D^{ret}(x-y)\bar{J}(y)dy,
$$
where
$$
D^{ret}_{\mu\nu}(x)=(g_{\mu\nu}-(1-\alpha )\frac
{\partial_{\mu}\partial_{\nu}}{\Box})D^{ret}(x),
$$
and $g_{\mu\nu}$ is the Minkowski metric. Free in-fields
satisfy
$$
(\Box g_{\mu\nu}-(1-\frac{1}{\alpha} )
\partial_{\mu}\partial_{\nu})A_{\nu}^{(in)}(x)=0,
$$
$$
\Box  c^{(in)}(x)=0,~ ~\Box \bar{c}^{(in)}(x)=0.
$$
and they are quantized in the Fock space with vacuum $|0>$.
The vector field $A_{\mu}^{(in)}$ is a Bose field and
the ghost fields $c^{(in)},\bar{c}^{(in)}$ are Fermi
fields. Actually one assumes a gauge $\alpha =1$. In a different
gauge one has to introduce additional ghost fields.
We introduce the notation $\psi_{i}=(A_{\mu},c,\bar {c})$
for the multiplet of  gauge and ghost fields.
The $U(N)$-invariant
Wightman functions  are defined as
\begin {equation} 
                                                          \label {3.4}
W(x_{1},...,x_{k})=
\frac{1}{N^{1+\frac{k}{2}}}<0|tr(\psi_{i_{1}}(x_{1})...
\psi_{i_{k}}(x_{k}))|0>.
\end   {equation} 
We will show that the limit of functions (\ref {3.4}) when
$N\to \infty$ can be expressed in terms of the master fields.
The master field for the gauge field $A_{\mu}(x)$ we denote
$B_{\mu}(x)$ and the master fields for the ghost fields
$c(x),\bar {c}(x)$ will be denoted $\eta(x),\bar {\eta}(x)$ .
The master fields  satisfy to equations
$$
D_{\mu} {\cal F}_{\mu \nu} +
\frac{1}{\alpha}\partial_{\nu}\partial_{\mu}B_{\mu}+
g\partial_{\nu}\bar{\eta} \eta +g\eta\partial_{\nu}\bar{\eta}=0,
$$
\begin {equation} 
                                                          \label {3.5'}
\partial_{\mu}(D_{\mu}\eta)=0,~~~
D_{\mu}(\partial_{\mu}\bar{\eta})=0
\end   {equation} 
where
\begin {equation} 
                                                          \label {3.0}
{\cal F}_{\mu \nu}=\partial_{\mu}B_{\nu}-\partial_{\nu}B_{\mu}
+g[B_{\mu},B_{\nu}],~ ~  ~
D_{\mu}\eta=\partial_{\mu}\eta+g[B_{\mu},\eta].
\end   {equation} 
These equations have the form of the Yang-Mills equations
 (\ref {3.2}) however the master fields
$B_{\mu}$, $\eta$, $\bar {\eta}$ do not have  matrix indexes
and they do not take values in a finite dimensional Lie algebra.
The gauge group for the field $B_{\mu}$ is an infinite dimensional group
of unitary operators in the Boltzmannian Fock space.
Equations (\ref {3.5'}) in terms of currents read
\begin {equation} 
                                                          \label {3.5}
\Box   B_{\nu} -(1-  \frac{1}{\alpha})\partial_{\nu}
\partial_{\mu}A_{\mu}=j_{\nu},
\end   {equation} 
$$
\Box \eta=j,~ ~~\Box \bar{\eta}=\bar{j},
$$
where
$$
j_{\nu} = -g\partial_{\mu}[B_{\mu},B_{\nu}]- g[B_{\mu},
{\cal F}_{\mu \nu}] - g\partial_{\nu}\bar{\eta}\eta
-g\eta\partial_{\nu}\bar{\eta},
$$
$$
j=-g \partial_{\mu}[B_{\mu},\eta],~~ ~\bar{j}=
- g[B_{\mu},\partial_{\mu}\bar{\eta}].
$$

We define the master fields by using the Yang-Feldman equations
\begin {equation} 
                                                          \label {3.6}
B_{\mu}(x)=B_{\mu}^{(in)}(x)+  \int D^{ret}_{\mu\nu}
(x-y)j_{\nu}(y)dy,
\end   {equation} 
$$
\eta(x)=\eta^{(in)}(x)+  \int D^{ret}(x-y)j(y)dy,
~ ~\bar{\eta}(x)=\bar{\eta}^{(in)}(x)+
\int D^{ret}(x-y)\bar{j}(y)dy,
$$
The in-master fields are quantized in the Boltzmannian Fock space. For
the master gauge field we have
\begin {equation} 
                                                          \label {3.7}
B_{\mu}^{(in)}(x)=\frac{1}{(2\pi)^{3/2}}\int \frac{d^{3}k}
{\sqrt{2|k|}}\sum _{\lambda =1}^{4}\epsilon ^{(\lambda)}
_{\mu}(k)[a^{(\lambda )*}(k)e^{ikx}+a^{(\lambda )}(k)e^{-ikx}) ,
\end   {equation} 
where $\epsilon ^{(\lambda)}_{\mu}(k)$  are polarization vectors
and annihilation and creation operators satisfy
\begin {equation} 
                                                          \label {3.8}
a^{(\lambda )}(k)a^{(\lambda ')*}(k')=
g^{\lambda \lambda '}\delta ^{(3)}(k-k'),
\end   {equation} 
The expression (\ref {3.7}) for the field $B_{\mu}(x)$ looks like
an expression for the photon field. However because of relations
(\ref {3.8})  the commutator  $[B_{\mu}(x),B_{\nu}(x)]$   does not
vanish and it  permits us to develop a gauge theory for the field
$B_{\mu}(x)$ with a non-Abelian gauge symmetry.

We quantize the master ghost fields  in the Boltzmannian Fock space
with indefinite metric
\begin {equation} 
                                                          \label {3.10}
\eta^{(in)}(x)=\frac{1}{(2\pi)^{3/2}}\int \frac{d^{3}k}
{\sqrt{2|k|}}(\gamma^*(k)e^{ikx}+\beta (k)e^{-ikx}) ,
\end   {equation} 
$$
\bar{\eta}^{(in)}(x)=\frac{1}{(2\pi)^{3/2}}\int \frac{d^{3}k}
{\sqrt{2|k|}}(\beta ^*(k)e^{ikx}+\gamma(k)e^{-ikx}) ,
$$
where creation and annihilation operators satisfy
$$
\gamma (k)\gamma ^{*}(k')=\delta ^{(3)}(k-k'),
$$
\begin {equation} 
                                                          \label {3.12}
\beta (k)\beta ^{*}(k')=-\delta ^{(3)}(k-k').
\end   {equation} 
We also assume that the product of any annihilation operator
with a creation operator of a different type
always is equal to zero, i.e.
$$
\gamma (k) \beta ^{*}(k')=\beta (k)\gamma ^{*}(k')=
a^{(\lambda )}(k)\gamma ^{*}(k)=0,$$
\begin {equation} 
                                                          \label {3.14}
a^{(\lambda )}(k)\beta ^*(k')=
\gamma (k)a^{(\lambda )*}(k')=\beta (k)a^{(\lambda )*}(k')=0.
\end   {equation} 
The Boltzmannian Fock vacuum satisfies
\begin {equation} 
                                                          \label {3.15}
\gamma (k)|\Omega _{0})=\beta (k)|\Omega _{0})=
a^{(\lambda )}(k)|\Omega _{0})=0.
\end   {equation} 
Let us denote $\chi _{i}=(B_{\mu}, \eta, \bar{\eta})$
the multiplet of the
master fields. The following theorem is true.

{\bf Theorem 2.} {\it At every order of perturbation
theory in the
coupling constant one has the following relation
\begin {equation} 
                                                          \label {3.16}
\lim _{N \to \infty}
\frac{1}{N^{1+\frac{k}{2}}}<0|\tr(\psi _{i_{1}}(x_1)...
\psi _{i_{k}}(x_k))|0>
=(\Omega_{0}|\chi _{i_{1}} (x_{1})...
\chi _{i_{k}}(x_{k})|\Omega_{0})
\end   {equation} 
where   the fields $A_{\mu} (x), c(x)$  and
$\bar {c}(x)$ are defined by
(\ref {3.3}) and $B_{\mu} (x), \eta(x)$  and $\bar {\eta}(x)$
are defined by  (\ref {3.6})}.

The proof of Theorem 2 is analogous to the proof of Theorem 1.
We get relations (\ref {3.12}) for master fields by taking
into account the wrong statistics of the ghost fields.

\section{Master Field for HP Diagrams}
\setcounter{equation}{0}
\subsection{Half-Planar Approximation for the One Mat\-rix Model}
A free $n$-point Green's function is defined as the vacuum expectation
of $n$-th power of master field
\begin{eqnarray}
\label{4}
G^{(0)}_n=\langle 0|\phi ^n |0\rangle .
\end{eqnarray}
As it is well-known, the Green's function (\ref{4}) is  given by a $n$-th
moment of Wigner's distribution  \cite{BIPZ,Voi}
\begin{eqnarray}
G^{(0)}_{2n}=\int_{-2}^{2}\frac{d \lambda }{2 \pi }
\lambda ^{2n}\sqrt{4-\lambda ^2}  =
\frac{(2n)!}{n!(n+1)!}.
\nonumber
\end{eqnarray}
This representation
can be also obtained as a solution of the Schwinger-Dyson equations
\begin{eqnarray}
G^{(0)}_{2n}=\sum_{m=1}^{n}G^{(0)}_{2m-2}G^{(0)}_{2n-2m}.
\nonumber
\end{eqnarray}

Interacting Boltzmann correlation functions are defined by the
formula \cite{AAV}
\begin{eqnarray}
\label{16}
G_n=\langle 0|\phi ^n
(1+S_{int}(\phi ))^{-1}
|0\rangle .
\end{eqnarray}
In contrast to the ordinary quantum field theory
where one deals with the exponential function of an interaction,
here we deal with  the
 rational function of an interaction.
In \cite{AZ}  it was shown that under natural assumptions  the form
(\ref{16})  is unique one
which admits  Schwinger-Dyson-like equations.

For the case of quartic interaction $S_{int}=g\phi ^4$
the Boltzmannian Schwinger-Dyson equations have the form
\begin{equation}
G_n=
\sum _{l=1}^{k-1}
G^{(0)}_{k-l-1}G_{l+n-k-1}+
\sum _{l=k+1}^{n}G^{(0)}_{l-k-1}G_{n+k-l-1}
\label{23}
\end{equation}
$$
-g[G_{n-k}G_{k+2}+
G_{n-k+1}G_{k+1}+
G_{n-k+2}G_{k}+
G_{n-k+3}G_{+k-1}].
$$
For 2- and 4-point correlation functions we have

\begin {equation} 
                                                          \label {24}
G_2=1-gG_2G_2-gG_4,
~~~G_4=2G_2-2gG_2G_4.
\end   {equation} 
\subsection{Boltzmann Correlation Functions for $D$-Dimensional $~~~~~$
Space-Time}

Here we  present the Schwinger-Dyson equations for Boltzmann
correlation functions in $D$-dimensional Euclidean space.
To avoid  problems with tadpoles let us following  \cite {AZsm}
consider the two-field formulation.
We adopt the following notations. Let
$\psi (x)=\psi ^+(x)+\psi ^-(x),$
$\phi (x)=$$\phi ^+(x)+$$\phi ^-(x)$
be the Bolzmann fields with creation and annihilation operators
satisfying the relations
\begin{eqnarray}
\psi ^-(x) \psi ^+(y)=
\phi ^-(x) \phi ^+(y)=D(x,y),
\nonumber
\end{eqnarray}
\begin{eqnarray}
\psi ^-(x) \phi ^+(y)=\phi ^-(x) \psi ^+(y)=0,
\nonumber
\end{eqnarray}
where
$
D(x,y)$
is $D$-dimensional Euclidean  propagator.
The $n$-point Green's function is defined by
\begin {eqnarray} 
\label {5.0}
 F_{n}(x_1,...,x_n)=
\langle 0|\psi (x_1)\phi (x_2)...
\phi (x_{n-1})\psi (x_n)
 (1+\int d^{D}x g \psi :\phi \phi : \psi))^{-1}
|0\rangle .
\end   {eqnarray} 

We define an one-particle irreducible  (1PI) 4-point function
$\Gamma_{4}(x,y,z,t)$  as
\begin {eqnarray} 
                              \label {ppp}
\Gamma_{4}(x,y,z,t)=
\int
dx^{\prime}dy^{\prime}dz^{\prime}
dt^{\prime}
F^{-1}_{2}(x,x^{\prime})
D^{-1}(y,y^{\prime})
D^{-1}(z,z^{\prime})\times
\end{eqnarray} 
$$
F^{-1}_{2}(t,t^{\prime}){\cal F}_{4}(x^{\prime},
y^{\prime},z^{\prime},t^{\prime}),$$
where
${\cal F}_{4} $ is a connected
part of $F_{4}$
$$F_{4}(x,y,z,t)={\cal F}_{4}(x,y,z,t)+
F_{2}(x,t)D(y,z).~~~~$$

Note that in the contrast to the usual case in the RHS of
(\ref {ppp}) we multiply ${\cal F}_{4}$ only on two full 2-point
Green functions while in the usual case to get an 1PI Green function one
multiplies  an $n$-point
Green function on $n$ full 2-point functions.

Let us write down  the Schwinger-Dyson equations
for the two- and four-point correlation functions.
We have
\begin {equation} 
\Gamma_{4}(p,k,r)= -g-
g\int dk^{\prime}
F_{2}(p+k-k^{\prime})D(k^{\prime})\Gamma_{4}(p+k-k^{\prime},
k^{\prime},r)
							  \label {bsl}
\end   {equation} 
\begin {equation} 
\label {si4}
\Sigma _{2} (p)=g\int dkdq F_{2}(k)D(q)
D(p-k-q)\Gamma _{4}(p,k,q).
\end   {equation} 
where
$$F_{2}=\frac{1}{p^{2}+m^{2}+\Sigma _{2}}
$$
Equation (\ref {bsl}) is the Bethe-Salpeter-like equ\-ation
with the kernel which contains an unknown  function $F_{2}$.
Equation (\ref {si4}) is similar to the  usual
relation between the self-energy function $\Sigma _{2}$ and the
4-point vertex function for $\varphi ^{4}$ field theory,
meanwhile equation (\ref{bsl}) is specific for the
Boltzmann field theory.
A special approximation reduces this system  of integral equations
to a linear integral equation which was considered \cite{Rothe}
in the rainbow approximation in the usual field theory.

\subsection{A Matreoshka of 2-particles Reducible Diagrams}
In  \cite {AAV} has been shown that equations sum up HP diagrams of
planar theory. Let us remind the definition of HP diagrams.
Sometimes they are called the rainbow diagrams.
The free rainbow diagrams are dual to tree diagrams and they
have been summed up in the zero dimensions  \cite {Bul}.
 The half-planar diagrams for  $<\tr (M(x_{n}),...M(x_{n}))>$ are defined
as a part of planar non-vacuum diagrams which are topologicaly equivalent
to the graphs with all vertexes lying on some
plane line in the left of generalized vertex
represented $\tr (M(x_{n}),...M(x_{n}))$ and all propagators lying in the
half plane.

We can  use $F_{2}$ and $F_{4}$ to construct correlations functions
which correspond the sum of more complicate diagrams.
Let us consider the following correlations functions
\begin {eqnarray} 
\label {5.01}
 F^{(1)}_{n}(x_1,...,x_n)=
\langle 0|\psi ^{(1)}(x_1)\phi (x_2)...
\phi (x_{n-1})\psi ^{(1)}(x_n)
 \end   {eqnarray} 
$$
(1+\int \prod _{i=1}^{i=4} d^{D}x_{i} \Gamma _{4}(x_{2},x_{3},x_{4},x_{1})
\psi ^{1} (x_{1}):\phi (x_{2})\phi (x_{3}): \psi ^{1} (x_{4})))^{-1}
|0\rangle ,
$$
here
\begin {equation} 
                                                          \label {ps1}
\psi ^{(1)-}(x) \psi ^{(1)+}(y)= F_{2}(x,y)
\end   {equation} 
The  Schwinger-Dyson equations
for the 2- and 4-point correlation functions
$ F^{(1)}_{2}$  and $ F^{(1)}_{4}$  satisfy to equations similar to equations
(\ref {bsl})  and (\ref {si4}). The obtained
$ F^{(1)}_{2}$  and $ F^{(1)}_{4}$ may be used to define the next
approximation to planar diagrams. One can see that such
procedure sums up special type of 2-particles reducible diagrams.
These diagrams are specified by the property that they contain  two
lines so that
after removing these lines from the given diagrams one reminds
with two disconnected parts and each of these disconnected part is itself
a connected 2-particle reducible diagram.  It is natural to call this set
 as an matreoshka of 2-particle reducible diagrams.

\subsection{Master Field for HP Gauge Theory}

In the case of gauge theory the set of the HP master field
is given by the field satisfying the following relations
\begin {equation} 
                                                          \label {6}
A^{-}_{\mu}(x)A^{+}_{\mu}(y)=\frac{1}{(2\pi)^{D}}\int d^{D} k
(g_{\mu \nu}-(1-\alpha)\frac{k_{\mu}k_{\nu}}{k^{2}})
\frac{1}{k^{2}}\exp ik(x-y),
\end   {equation} 

\begin {equation} 
                                                          \label {7}
\bar {c}^{-}(x)c^{+}(y)=\frac{1}{(2\pi)^{D}}\int d^{D} k
\frac{1}{k^{2}}\exp ik(x-y),
\end   {equation} 

\begin {equation} 
                                                          \label {8}
c^{-}(x)\bar {c}^{+}(y)=-\frac{1}{(2\pi)^{D}}\int d^{D} k
\frac{1}{k^{2}}\exp ik(x-y),
\end   {equation} 

\begin {equation} 
                                                          \label {9}
A^{-}_{\mu}(x)c^{+}(y)=A^{-}_{\mu}(x)\bar {c}^{+}(y)=0
\end   {equation} 

A unrenormalized
interacting Lagrangian which assumed to enter in the correlation functions as
is (\ref {5.0}) (a generalization to two-fields formalism is evident)
has the form
$$L_{unr}=\frac{g}{4}\{~\partial _{\nu}A_{\mu}[A_{\mu}, A_{\nu}]
+A_{\nu}\partial _{\nu}A_{\mu}A_{\mu}
-A_{\mu}\partial _{\nu}A_{\mu}A_{\nu}
+[A_{\nu},A_{\mu}]\partial _{\nu}A_{\mu} \}+
$$
$$4g^{2}\{2A_{\nu}A_{\mu}A_{\nu}A_{\mu}-
A_{\nu}A_{\mu}A_{\mu}A_{\nu}-A_{\mu}A_{\mu}A_{\nu}A_{\nu}\}$$
\begin {equation} 
                                                          \label {5}
-g\{\partial _{\mu}\bar {c} A_{\mu}c
+c\partial _{\mu}\bar {c} A_{\mu}+ A_{\mu}c\partial _{\mu}\bar {c}
-\partial _{\mu}\bar {c}c A_{\mu}
-A_{\mu}\partial _{\mu}\bar {c}c
-cA_{\mu}\partial _{\mu}\bar {c}\}
\end   {equation} 
Divergences in one-loop correlation functions may be removed
by the following renormalizations
$$L_{r}=\frac{g}{4}Z_{1}^{HP}(Z_{2}^{HP})^{-1}
\{\partial _{\nu}A_{\mu}[A_{\mu}, A_{\nu}]+
+A_{\nu}\partial _{\nu}A_{\mu}A_{\mu}
-A_{\mu}\partial _{\nu}A_{\mu}A_{\nu}
+[A_{\nu},A_{mu}]\partial _{\nu}A_{\mu} \}+
$$
$$+\frac{g^{2}}{2}Z_{4}^{HP}(Z_{2}^{HP})^{-1}\{2A_{\nu}A_{mu}A_{\nu}A_{mu}-
A_{\nu}A_{mu}A_{\mu}A_{\nu}-A_{\mu}A_{\mu}A_{\nu}A_{\nu}\}
-$$
$$\frac{g}{8}\tilde {Z}_{1}^{HP}
(\tilde {Z}_{2}^{HP})^{-1}\partial _{\mu}\bar {c} A_{\mu}c
-cA_{\mu}\partial _{\mu}\bar {c})$$
\begin {equation} 
                                                          \label {10}
+\frac{g}{8}\tilde {Z}_{1}^{HP}(Z_{2}^{HP}\tilde {Z}_{2}^{HP})^{-1/2}
(c\partial _{\mu}\bar {c} A_{\mu}+
A_{\mu}c\partial _{\mu}\bar {c}
-\partial _{\mu}\bar {c}c A_{\mu}
-A_{\mu}\partial _{\mu}\bar {c}c)
\end   {equation} 
where
\begin {equation} 
                                                          \label {12}
Z_{1}^{HP}=1+\frac{g^{2}}{16\pi ^{2}}(-\alpha)\ln \Lambda
\end   {equation} 
\begin {equation} 
                                                          \label {13}
Z_{2}^{HP}=1+\frac{g^{2}}{16\pi ^{2}}(\frac{13}{3}-\alpha)\ln \Lambda
\end   {equation} 

\begin {equation} 
                                                          \label {15}
\tilde {Z}_{1}^{HP}=1,~~
\tilde {Z}_{2}^{HP}=
1+\frac{g^{2}}{16\pi ^{2}}(\frac{3}{2}-\frac{\alpha}{2})\ln \Lambda
\end   {equation} 
$Z_{2}$ factors enter in the nonstandard way since only two
legs (the first and the last) may bring the wave function renormalization.
So we have a modification in the definition of beta function
\begin {equation} 
                                                          \label {11}
\beta ^{HP}=-g\frac{\partial }{\partial (\ln \Lambda)}
(\frac{Z^{HP}_{2}}{Z^{HP}_{1}})
\end   {equation} 
Therefore we have
\begin {equation} 
                                                          \label {14}
\beta ^{HP}=-\frac{g^{2}}{16\pi ^{2}}(\frac{13}{3}-\alpha
+\alpha)=-\frac{g^{2}}{16\pi ^{2}}\frac{13}{3}
\end   {equation} 
Recall that the usual beta function is
$\beta =-\frac{g^{2}}{16\pi ^{2}}\frac{11}{3}$.
Note that we get  good results:
beta remains negative and in this approximation it does not depend
on  the gauge fixing parameter $\alpha$, i.e. it is gauge invariant.

\section   {Concluding remarks}

In conclusion,  models of quantum field theory with interaction
in the Boltzmannian Fock space have been considered.

To define the master field for large N matrix models
we used the Yang-Feldman equation with a free field
quantized in the Boltzmannian Fock space. The master field for gauge theory
does not take values in a finite-dimensional Lie algebra however there is a
non-Abelian gauge symmetry. For the construction of the master field it was
essential to work in Minkowski space-time and to use the Wightman correlation
functions.
The fact that the master field
satisfies the same equations as usual relativistic fields push us to
develop a non-perturbative methods of investigation of these equations.
Note in this context that in all previous attempts
of approximated treatment of planar theory were used some non-perturbative
approximation \cite {Sl,IA,Fer}.

To sum up a part of planar diagrams we have used the new interaction
representation with a rational function of the
interaction Lagrangian instead of
the exponential function in the standard interaction representation.
The Schwinger-Dyson equations  for the 2- and  4-point correlation
functions for this theory form a closed system of equations.
The solutions of these
equations may be used to sum up a more rich class of planar diagrams.
This is a subject of further investigations.
$$~$$

{\bf ACKNOWLEDGMENT}
$$~$$
The author is grateful to P.Medvedev, I.Volovich and A.Zubarev
for useful discussions.

\end{document}